**Title:** Mapping PM$_{2.5}$ concentration at sub-km level resolution: a dual-scale retrieval method

**Authors:** Qianqian Yang [a], Qiangqiang Yuan [*][a,b], Linwei Yue [*][c], Huanfeng Shen [d], Liangpei Zhang [e]

**Affiliations:**

[a] School of Geodesy and Geomatics, Wuhan University, Wuhan, Hubei, 430079, China

[b] Key Laboratory of Geospace Environment and Geodesy, Ministry of Education, Wuhan University, Wuhan 430079, Hubei, China

[c] Faculty of Information Engineering, China University of Geosciences, Wuhan, Hubei, 430074, China

[d] School of Resource and Environmental Sciences, Wuhan University, Wuhan, Hubei, 430079, China

[e] State Key Laboratory of Information Engineering, Survey Mapping and Remote Sensing, Wuhan University, Wuhan, Hubei, 430079, China

[*] **Corresponding authors:**

Qiangqiang Yuan (yqiang86@gmail.com)
Linwei Yue (yuelw@cug.edu.cn)

# ABSTRACT

Satellite-based retrieval has become a popular PM2.5 monitoring method currently. To improve the retrieval performance, multiple variables are usually introduced as auxiliary variable in addition to aerosol optical depth (AOD). Different kinds of variables are usually at different resolutions varying from sub-kilometers to dozens of kilometers. Generally, when doing the retrieval, variables at different resolutions are resampled to the same resolution as the AOD product to keep the scale consistency. A deficiency of doing this is that the information contained in the scale difference is discarded. To fully utilize the information contained at different scales, a dual-scale retrieval method is proposed in this study. At the first stage, variables which influence PM2.5 concentration at large scale were used for PM2.5 retrieval at coarse resolution. Then at the second stage, variables which affect PM2.5 distribution in finer scale, were used for the further PM2.5 retrieval at high resolution (sub-km level resolution) with the retrieved PM2.5 at the first stage at coarser resolution also as input. In this study, four different retrieval models including multiple linear regression (MLR), geographically weighted regression (GWR), random forest (RF) and generalized regression neural network (GRNN) are adopted to test the performance of the dual-scale retrieval method. Compared with the traditional retrieval method, the proposed dual-scale retrieval method can achieve PM2.5 mapping at finer resolution and with higher accuracy. Dual-scale retrieval can fully utilize the information contained at different scales, thus achieving a higher resolution and accuracy. It can be used for the generation of quantitative remote sensing products in various fields, and promote the improvement of the quality of quantitative remote sensing products.

# 1. Introduction

Fine particulate matter with an aerodynamic diameter of less than 2.5 μm (PM$_{2.5}$) holds a great threat to ecological environment and public health(Ho et al., 2018). Ground environmental monitoring sites have been built in the worldwide for the measuring of PM2.5 concentration. However, sites-based measurement cannot achieve the monitoring in large extent with continuous spatial coverage(Shen et al., 2018). Therefore, satellite based remote sensing retrieval method has become one of the mainstream methods for PM2.5 monitoring in recent years(Li et al., 2017a; Li et al., 2017b).

The basic satellite product required for PM2.5 retrieval is the aerosol optical depth (AOD), which usually holds a spatial resolution at kilometers level, for instance, 10km (MOD04), 3km, (MOD04_3K), and 1km (MAIAC) (Liu et al., 2019). In addition to AOD product, other variables such as meteorological and topographical factors are also included as auxiliary variables, to promote the performance of the retrieval model(Bi et al., 2019; Chen et al., 2018). Meteorological variables are usually at a coarse resolution (dozens of kilometers level), while topographical data (such as DEM and landcover) are usually at fine resolution (sun-kilometers level). We can notice that the resolution of variables used for PM2.5 retrieval usually varies in a wide range, from sub-kilometers to dozens of kilometers. When establishing the retrieval model, to keep the scale consistency, the input variables mentioned above are usually resampled to the same resolution as the used AOD product (we call this "single-scale retrieval")(Boys et al., 2014; Li et al., 2017a). To be specific, variables at higher resolution than AOD product are downsampled, and variables at lower resolution than AOD product are upsampled, both to the resolution of AOD product. However, variables at different resolution often contain information at different levels of detail. Simple resample to the same resolution may result in the information loss, especially the loss of detail information at sub-kilometers level.

Therefore, we propose the concept of dual-scale retrieval, which retrieves PM2.5 concertation through two stages. At the first stage, PM2.5 concentration at a coarse resolution are retrieved with variables at resolution lower than or equal to AOD's resolution. Then at the second stage, the retrieved low-resolution PM2.5 concentration together with variables at higher resolution than AOD are used for PM2.5 retrieval at fine scale. The two-stage dual-scale retrieval method can make fuller use of the information embodied in the scale difference, and therefore has the potential to bring about the improvement in both product resolution and model performance.

In this study, we selected two linear retrieval model, i.e. multiple linear regression (MLR) (Xu et al., 2018b) and geographically weighted regression (GWR) (Jiang et al., 2017) and two machine-learning based nonlinear model, i.e. random forest (RF) (Hu et al., 2017; Huang et al., 2018) and generalized regression neural network (GRNN) (Li et al., 2017b) to test the performance of the dual-scale retrieval method. For a fair comparison with the single-scale retrieval method, we used the same model in the two steps. Meteorological factors which include nine variables and MODIS AOD product at 0.1° resolution were used for the first-scale retrieval, DEM and landcover data were then used for the second-scale estimation. 10-fold cross validation and dense point cross validation were used for the quantitation evaluation of the model performance. After the model building, we mapped the annual PM2.5 concentration for five typical cities in China at the 0.003°×0.003° resolution and the spatial distribution feature of PM2.5 concentration was analyzed. The quantitative evaluation and mapping results showed that the dual-scale method can not only achieve better model performance with high retrieval accuracy; but also output the PM2.5 product with higher resolution and capture the fine-scale spatial variations better than the traditional single-scale retrieval method.

Furthermore, the proposed dual-scale retrieval method can not only be used for PM2.5 concentration mapping, but also has great potential in the production of other quantitative remote sensing products, such as soil moisture (Xu et al., 2018a) and some vegetation parameters (Yuan et al., 2019). With both the accuracy and resolution improved, the application value of quantitative remote sensing products can be greatly improved. To sum up, the dual-scale retrieval method considers the information contained in the scale differences, and make a fuller use of it through a two-stage different-scale retrieval. The better extraction of the information contained in the various scales then brings about the improvement in both product resolution and model performance.

The rest of this paper is organized as follows. Section 2 is the data and method part, where we introduce the data sources, and the methodology, and provide a flow chart of the study design. The experimental results and a discussion are provided in Section 3. Finally, we make a summary of our work in Section 4.

**2. Data and method**

*2.1. Study domain*

China is a large country with dense population and broad territory. The rapid economic development in recent decades have brought serious pollution in China(Ma et al., 2016). Since 2013, multiple environmental monitoring sites have been built in China for the monitoring of air quality, providing a foundation for the air pollution research. In the study, PM2.5 data from these stations are used for model building. As for the retrieval phase, we chose five typical cities in China for the mapping of PM2.5 concentration. The selected cities include Beijing, Wuhan, Shanghai, Guangzhou and Chengdu. Beijing, the capital city of China, locates in the North China Plain. Beijing is marked by its flatness and arid climate. There are only three hills to be found in the city limits and mountains surround the capital on three sides. Affected by the fast urbanization in recent years and unfavorable topography condition, PM2.5 pollution has become an urgent problem for Beijing (Guo et al., 2017). Wuhan is the largest cities in Central China with a dense pollution of 11,081,000. With the Yangtze River runs through the city, Wuhan has a humid climate and plain is the dominant terrain. Impacted by the heavy industry production, the PM2.5 pollution is also a grave problem for Wuhan (Zhang et al., 2018). Shanghai, the largest economic and transportation center in China, located in the Yangtze River Delta in East China and sits on the south edge of the mouth of the Yangtze in the middle portion of the eastern Chinese coast. As the cradle of China's modern industry, Shanghai has undertaken much industrial production in China. Combined with the fast economic development, PM2.5 concentration has also been increasing in recent decades (Xiao et al., 2017). Guangzhou is the central city of South China and is located at the flourish Pearl River Delta region. With the Tropic of Cancer crossing through north of the city and Pearl River flowing across the city, Guangzhou enjoys favorable weather which is warm and humid. Compared with the aforementioned cities, PM2.5 pollution in Guangzhou is less serious(Yang et al., 2017b). Chengdu is located at the western edge of Sichuan Basin and sits on the Chengdu Plain; the dominant terrain is plains but surround by high mountains. Chengdu has similar climate to Wuhan——enough precipitation , humid and mild. Chengdu is also one of the most important economic centers, transportation and communication hubs in Western China. The unique topography structure makes the air pollution in Chengdu special as well (Ning et al., 2018). The chosen cities cover different topographies, climates and pollution degrees for a comprehensive display of the model performance. And the locations and topography of these cities are displayed in Fig.1.

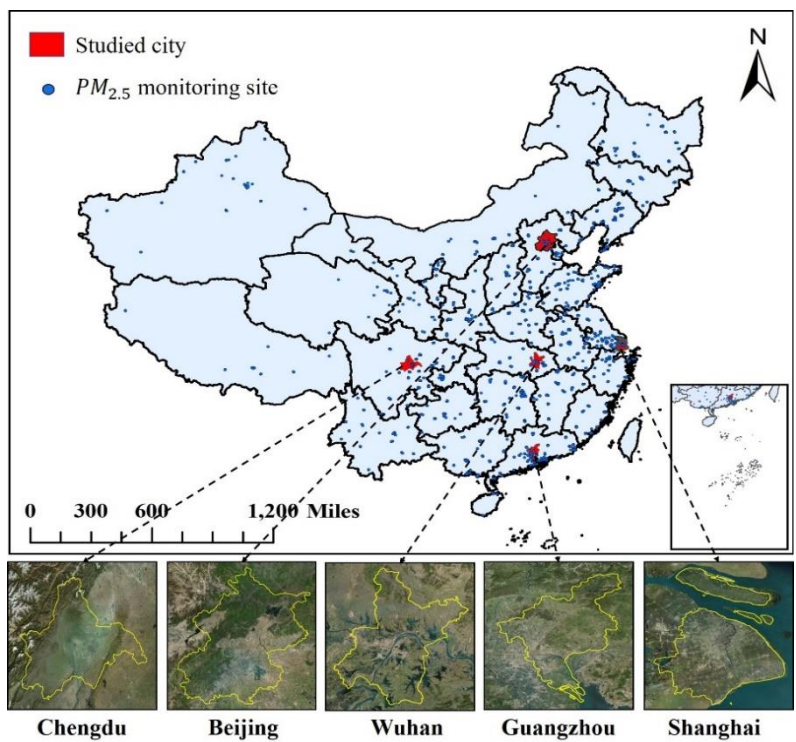

**Fig. 1** Study domain. The map above shows the distribution of PM2.5 monitoring sites (blue points) and the locations of the five studied cities (red polygons). The satellite images below show the topography of the five cities while yellow lines on them are administrative boundaries of cities. The image was provided by ArcGIS online world imagery.

*2.2. Dataset*

*2.2.1 Ground sites $PM_{2.5}$ data*

The ground-based PM2.5 concentration data from environmental monitoring stations was provided by the Ministry of Ecology and Environment of the People's Republic of China (http://www.mee.gov.cn/). Hourly PM2.5 concentration data in more than 1400 sites for 2015 was collected and then averaged to annual data after an outlier filtering. The distribution of the monitoring sites is shown in Fig.1.

*2.2.2 AOD data*

Aerosol optical thickness (AOD) represents the vertical integral of the aerosol extinction coefficient on the atmospheric column (Beloconi et al., 2016). It is an indirect measure of the particles present in the air, thus has been widely used for retrieving PM2.5 concentration (Li et al., 2017a; Li et al., 2017b; Shen et al., 2018; Yang et al., 2019). The satellite AOD product used in this study was downloaded from Level-1 and Atmosphere Archive and Distribution System (LAADS) Distributed Active Archive Center (DAAC) of NASA. The Moderate Resolution Imaging Spectroradiometer (MODIS) Collection 6 level 2 daily AOD data from the Terra (MOD04_L2), which are reported at 10 km (~0.1°) was used in this study (Levy et al., 2013; Levy et al., 2007).

*2.2.3 Meteorological data*

Many researchers have proved that meteorological conditions can have a significant impact on PM2.5 concentration(Yang et al., 2017a), and the introduction of meteorological factors can improve the retrieval accuracy. In this study, several commonly used meteorological variables including temperature (TMP), pressure (PS), relative humidity (RH), zonal wind speed (UWS), meridional wind speed (VWS), the lifted index (LI), vertical speed (VS), precipitation (PR) were considered in the retrieval model, and were obtained from the NCEP/NCAR Reanalysis 1 project. Besides, planetary boundary layer height (PBLH) were also important for PM2.5 estimation (Su et al., 2018; Wang et al., 2019b), we derived the PBLH data from the MERRA-2 dataset M2T1NXFLX collection.

*2.2.4 Topographical data*

Though not used as much as meteorological factors in PM2.5 retrieval problem, topographical factors can also effect PM2.5 pollution a lot and can help improve the performance of the retrieval model (Beloconi et al., 2016; Jung et al., 2017; Wang et al., 2017). Therefore, topographical factors, such as land cover (LC) and elevation (DEM) were also considered in this study. The land-cover product was provided by the European Space Agency (ESA) Climate Change Initiative (CCI), with a spatial resolution of 300m. The 30-m elevation data were obtained from the Global Multi-Data Fused Seamless DEM product (GSDEM-30)(Yue et al., 2015; Yue et al., 2017), and can be downloaded from http://sendimage.whu.edu.cn/res/DEM_share/. To keep consistency with land cover data, the 30-m DEM data was resampled to 300m through simple pixel average. So, the used topographical data, including DEM and landcover data, were both at a 300-m resolution.

*2.3. Methodology*

*2.3.1 Model development*

The process of the dual-scale retrieval method mainly includes two stages. In the first stage, a low-resolution PM2.5 product was retrieved with AOD and meteorological factors. Meteorological data is resampled to match the AOD grids, and the retrieved PM2.5 product is at the resolution of AOD. Considering that PM2.5 concentration, as a geographical variable, usually contains strong spatial autocorrelation (Li et al., 2017a), the longitude and latitude information was also input into the model. Hence, the process of stage one can be shown as:

$$PM_{2.5\_L} = f(lat, lon, AOD, TMP, PS, RH, UWS, VWS, LI, VS, PR, PBLH)$$

Where $PM_{2.5\_L}$ stands for the retrieved low resolution PM2.5 concertation, $f()$ represents the retrieval model, and in this study, i.e. MLR, GWR, RF, and GRNN model. In the second stage, the final high-resolution PM2.5 product was obtained with low-resolution PM2.5 product produced in the first stage, DEM and landcover product as input. To unify coordinate system (AOD and meteorological data are using a geographic coordinate system and topographical data using a projected coordinate system), the projected coordinate system was first reprojected into the World Geodetic System 1984 (WGS84) geographic coordinate system. Then we resampled the 300-m landcover and DEM data into 0.003° (10km~=0.1°), and generate the final high-resolution PM2.5 product at the resolution of 0.003°. The $PM_{2.5\_L}$ product from stage one was also resampled to the resolution of 0.003° using bilinear interpolation. Therefore, the upsampled low resolution (0.003°×0.003°) PM2.5 product from stage one, together with the resampled 0.003° landcover and DEM data, was input into the second-stage retrieval model. At a higher resolution, location information may have a different impact on PM2.5 distribution, so, the longitude and latitude was introduced into the model for the second time. This process can be simply written as:

$$PM_{2.5\_H} = f(lat, lon, PM_{2.5\_L}, LC, DEM)$$

Where $PM_{2.5\_H}$ stands for the final high resolution PM2.5 product, $f()$ represents the retrieval model, i.e. MLR, GWR, RF, and GRNN model.

For each stage, we first preprocess the data to get data pairs for model training. The preprocessing procedure include simple gaps filling and data match. AOD and DEM data missing in a small spatial range is filled with the inverse distance weighted (IDW) interpolation. Then all the raster data are matched with ground environmental stations according to longitude and latitude. Secondly, the obtained data pairs are used for training the retrieval model, with PM2.5 concentration as output and other data as input. The overall workflow is shown in Fig.2.

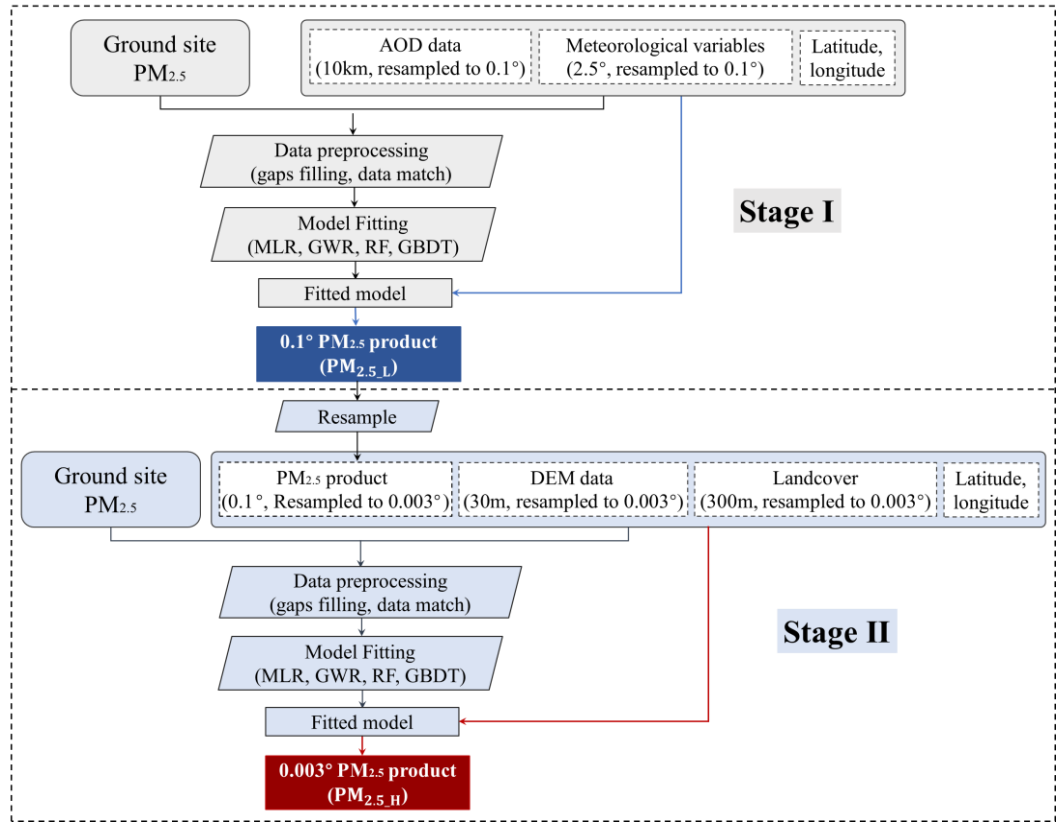

**Fig. 2** Flow chart of the dual-retrieval method. The upper grey part stands for the first stage and the lower light-blue part is the second stage.

To fully verify the performance of the dual-scale retrieval method, we selected the traditional single-scale retrieval method for a comparison, which can be expressed as:

$$PM_{2.5\_S} = f(lat, lon, AOD, TMP, PS, RH, UWS, VWS, LI, VS, PR, PBLH, LC, DEM)$$

Where $PM_{2.5\_S}$ stands for the PM2.5 retrieved from the single-scale method, $f()$ represents the retrieval model, which is the same as the dual-scale retrieval method. It should be noticed that the $PM_{2.5\_S}$ product is at the resolution of $0.1°\times 0.1°$, that's to say, nearly 30 times lower than the resolution of dual-scale retrieval results——$PM_{2.5\_H}$.

Four retrieval models, including two linear models and two nonlinear models, were selected for test. The two linear regression model were MLR and GWR. MLR fits an observed dependent data set (i.e., PM2.5 concentration) using a linear combination of independent variables (Kokaly and Clark, 1999). MLR has been widely used in remote sensing applications because of its simplicity, but it relies on several assumptions concerning data distributions, and its performance depends on meeting these assumptions as well as the linearity of the modeled relationship (Xu et al., 2018b). GWR is a development from the MLR model, which blends spatial heterogeneity into the regression model and build a spatially varying relationship between the studied variables. Being able to capture the spatial variations in local effects, GWR can usually achieve a better performance than MLR model, and has been a popular model for PM2.5 retrieval in recent decades (Hu et al., 2013; Wang et al., 2019a). However, both the models mentioned above are linear regression models, and may not be able to capture the complicated relationship between PM2.5 and the multifarious predictors at various resolution. And in recent years, nonlinear machine-learning (ML) based models have shown satisfied performance in PM2.5 retrieval problem, outperformed the linear models (Li et al., 2017a; Li et al., 2017b; Liu et al., 2017). Therefore, two ML-based methods are also tested in this study. The first is the RF model, RF is an ensemble-based decision tree approach, which consists of a combination of decision trees fitted by randomly selected subsets of training samples. Final predictions produced by RF model are determined by the average of the results of all the trees(Xu et al., 2018b). Another ML-based algorithm is the GRNN model. GRNN is a special form of a radial basis function neural network. Compared with the most common back-propagation neural network, it overcomes the disadvantages of slow convergence and easily convergence to local minima. Meanwhile, compared with the popular feedforward neural networks, the GRNN has the advantages of a relatively simple structure, rapid training, low computational cost, and global convergence (Yuan et al., 2019). Therefore, it has been used for retrieval problem and shows a good performance(Li et al., 2017b; Xu et al., 2018a).

*2.3.2 Model validation*

The 10-fold cross-validation technique was used to validate the performance of the proposed retrieval method. The dataset was averagely divided into 10 folds randomly. Nine folds of the dataset were used for model fitting, and the left one was predicted in each round of the cross-validation. This step was repeated 10 times until every fold was tested (Shen et al., 2018). And then the coefficient of determination ($R^2$) was calculated for the quantitatively indication of the model performance.

In addition to the commonly used cross-validation, we also conducted another validation method, which we called "dense point cross validation". As the finally generated product has a higher resolution than the general product, when several sites are located at the same pixel on the general product, these sites can correspond to different pixels on the high-resolution product. We call these ground sites "dense points". If the generated high-resolution product can keep high consistency with the PM2.5 value of these dense points, it means the generated detail information in the high-resolution map can well capture the real PM2.5 variations. This process can be explained by Fig.3. When conducting the 10-fold cross-validation, the Pearson correlation coefficients (r) between the PM2.5 concentration of these dense points and these corresponding red grids are calculated as indicator.

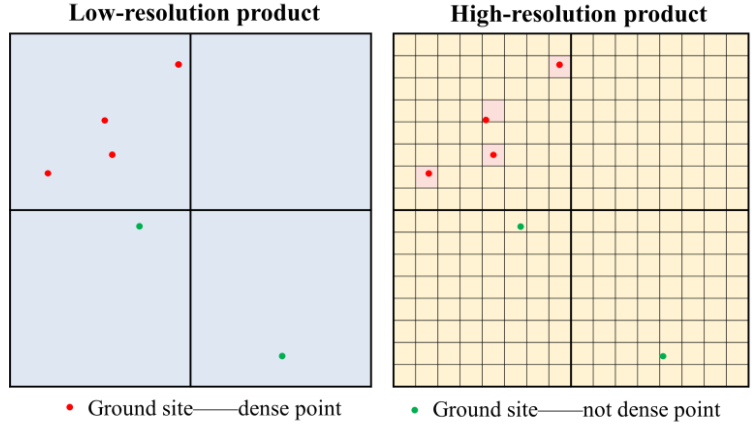

**Fig. 3** Schematic for dense points cross validation. The light blue square represents the low-resolution product and the light-yellow square represents the high-resolution product. Each grid stands for one pixel. Points stand for the ground sites while red points are the dense points. The pink grids are the corresponding pixels to dense points.

**3. Results and discussion**

*3.1. Model performance*

The results of model fitting and validation are shown in Table 1. We can find that, for most cases, the proposed dual-scale retrieval method has a significant improvement on performs compared to the traditional single-scale retrieval method. Specifically, the cross-validation $R^2$ improves 0.07, 0.05, and 0.05 for the GWR, RF and GRNN models respectively. This result shows that dual-scale retrieval can mine the information contained in the scale difference among the input variables better, and then the richer information brings about the promotion of the model performance. Among the four retrieval models, GWR shows the best performance, with the fitting $R^2$ and cross validation $R^2$ reaches 0.87 and 0.86. Followed by GWR, RF model also has a pretty good performance in terms of quantitative evaluation, the fitting $R^2$ is 0.80 and the cross-validation $R^2$ is 0.79. Then come GRNN and MLR. MLR model shows the worst performance among the four models with a cross-validation $R^2$ of 0.63. This proves that simple linear regression may not be able to well describe the complex relationship between PM2.5 and multiple influencing factors, and the scale differences of these variables at different resolution.

Table 1. The model fitting and cross-validation results. Fitting $R^2$ is the $R^2$ score for model fitting and CV $R^2$ represent the $R^2$ score for cross validation.

|  |  | MLR | | GWR | | RF | | GRNN | |
|---|---|---|---|---|---|---|---|---|---|
|  |  | Fitting $R^2$ | CV $R^2$ | Fitting $R^2$ | CV $R^2$ | Fitting $R^2$ | CV $R^2$ | Fitting $R^2$ | CV $R^2$ |
| **Dual-scale** | Stage I | 0.59 | 0.58 | 0.84 | 0.78 | 0.74 | 0.72 | 0.61 | 0.60 |
|  | Stage II | 0.64 | **0.63** | 0.87 | **0.86** | 0.80 | **0.79** | 0.69 | **0.68** |
| **Single-scale** |  | 0.64 | 0.62 | 0.87 | 0.79 | 0.76 | 0.74 | 0.64 | 0.63 |

We also show the scatter plots of the cross-validation results in Fig. 4. Apart from the highest cross-validation $R^2$ score, the fitting line of the scatters of GWR model is also the closest to the 1:1 line with the slope equaling to 0.88, indicating a small bias. In comparison, fault occurs in the scatter plot of RF, we think that indicates the estimated PM2.5 concentration of RF model is not continuous enough, which may bring some problems when mapping PM2.5 concentration. As for the MLR and GRNN model, no fault appears and the slope for fitting line are 0.62 and 0.46 respectively. Though the results of GRNN has a higher cross-validation R2 than MLR, but the slope is far from 1:1 line. That indicate GRNN can achieves a good linear correlation between estimated value and ground truth, but the estimated values contain a large bias. All the slope in Fig.4(a)-(d) are smaller than 1, which means overestimation for lightly polluted regions and underestimation for highly estimated regions still exist. This is also a common problem for $PM_{2.5}$ retrieval research (He and Huang, 2018; Xue et al., 2019).

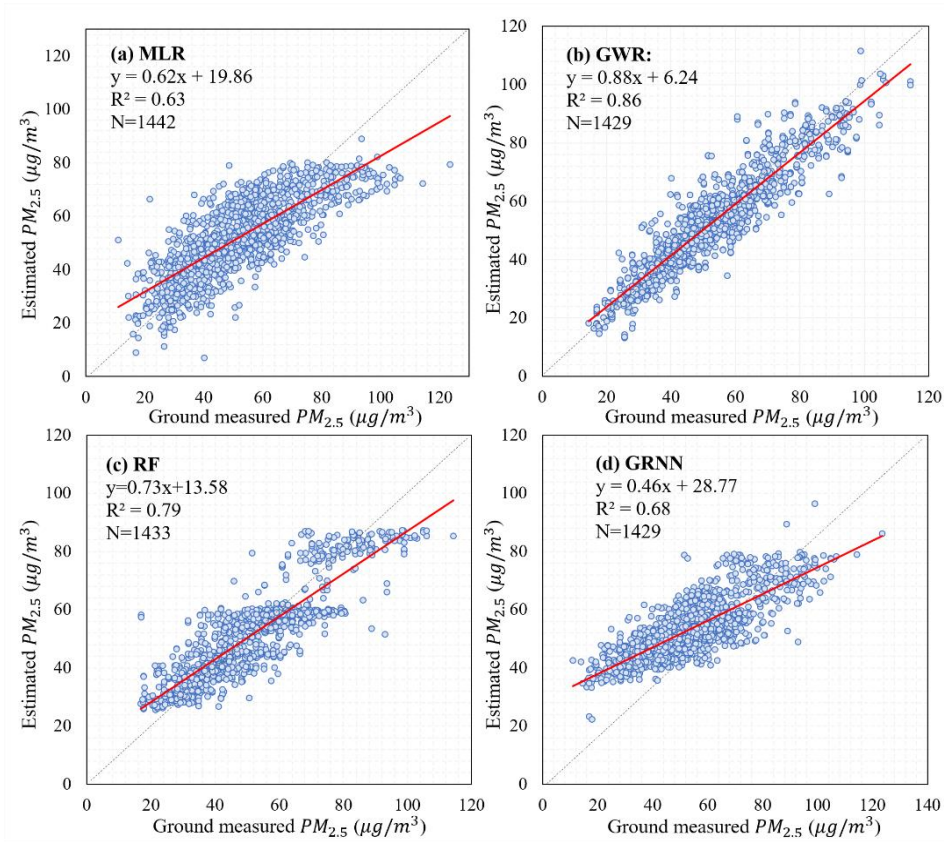

**Fig. 4** The scatter plots for the cross-validation results of four retrieval models. (a) MLR-based two scale retrieval; (b) GWR-based two scale retrieval; (c)RF-based two scale retrieval;(d) GRNN-based two scale retrieval. The grey dashed lines stand for the 1:1 line; and the red lines are the fitted lines for the scatters. N stands for the sample numbers.

The results of dense point cross validation are displayed in Fig. 5 and Fig. 6. Fig. 6 shows the specific process for dense point cross validation taking the retrieval results of Guangzhou as an example. The two pictures above are the results of the single-scale and dual-scale retrieval in Guangzhou. We can clearly see that in the results of single-scale retrieval, there are many sites which have different PM2.5 concentration values located in the same pixel (marked by black rectangles). The detailed spatial variations are covered by coarse pixels. On the contrary, the results of dual-scale retrieval then can build more detail information about spatial variations, and these sites can correspond to different pixels. For a quantitative evaluation of the accuracy of the detail information built by dual-scale retrieval method, we calculate the correlation between these dense points and corresponding pixels and the results are listed in the Fig.5. The correlation coefficient reaches 0.78, and for single-scale retrieval, the value is just 0.53. That means the detail information are built and correctly built in dual-scale retrieval.

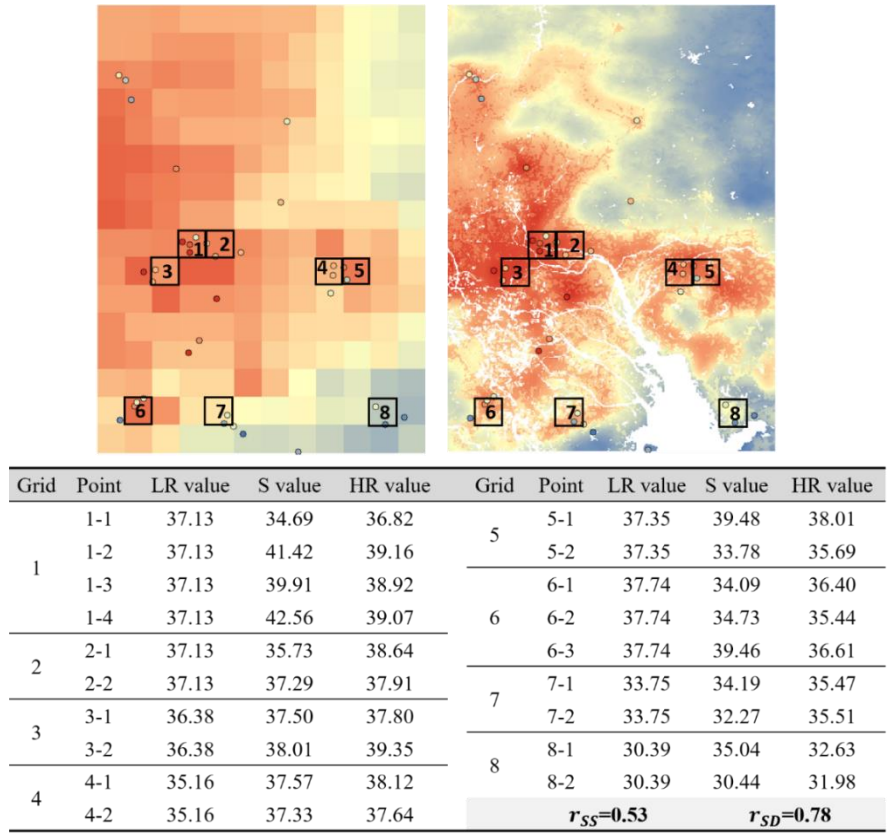

| Grid | Point | LR value | S value | HR value | Grid | Point | LR value | S value | HR value |
|---|---|---|---|---|---|---|---|---|---|
| 1 | 1-1 | 37.13 | 34.69 | 36.82 | 5 | 5-1 | 37.35 | 39.48 | 38.01 |
| | 1-2 | 37.13 | 41.42 | 39.16 | | 5-2 | 37.35 | 33.78 | 35.69 |
| | 1-3 | 37.13 | 39.91 | 38.92 | 6 | 6-1 | 37.74 | 34.09 | 36.40 |
| | 1-4 | 37.13 | 42.56 | 39.07 | | 6-2 | 37.74 | 34.73 | 35.44 |
| 2 | 2-1 | 37.13 | 35.73 | 38.64 | | 6-3 | 37.74 | 39.46 | 36.61 |
| | 2-2 | 37.13 | 37.29 | 37.91 | 7 | 7-1 | 33.75 | 34.19 | 35.47 |
| 3 | 3-1 | 36.38 | 37.50 | 37.80 | | 7-2 | 33.75 | 32.27 | 35.51 |
| | 3-2 | 36.38 | 38.01 | 39.35 | 8 | 8-1 | 30.39 | 35.04 | 32.63 |
| 4 | 4-1 | 35.16 | 37.57 | 38.12 | | 8-2 | 30.39 | 30.44 | 31.98 |
| | 4-2 | 35.16 | 37.33 | 37.64 | | $r_{SS}$=0.53 | | $r_{SD}$=0.78 | |

**Fig. 5** The dense point validation results: an example of GWR retrieved results in Guangzhou. Black rectangles are grids where dense points located. These grids are numbered from top to bottom, left to right. Point in the grids are also number by grid number-*, from top to bottom, left to right. LR and HR value are the estimated PM2.5 value of single-scale and dual-scale retrieval respectively, and S value is the ground sites PM2.5 value. $r_{SD}/r_{SS}$ stands for the Pearson correlation coefficient between ground site PM2.5 and dual-scale/single-scale retrieved PM2.5 concentration.

Fig. 6 (a)-(d) represents the dense point cross-validation results of MLR, GWR, RF and GRNN respectively for whole study area, where there are nearly 900 dense points. The predicted values at these sense points are calculated at each fold of cross validation, and the red and blue lines stand for the predicted value of dual-scale and single-scale respectively. And the dark line is the value of ground measured PM2.5 concentration, which is sorted from smallest to largest. Being closer to the dark line can indicate a better expression of the detail PM2.5 distribution. Fig.6 shows that the red lines, which stand for the results of dual-scale retrieval, are closer to the dark lines for most cases. That means the dual-scale retrieval method did capture the detail variations of PM2.5 concentration correctly. The correlation coefficients between the PM2.5 concentration at dense points and the single/dual-scale retrieval results are also calculated, which are referred as $r_{SS}$ and $r_{SD}$. For most cases, $r_{SD}$ is much large than $r_{SS}$. Specifically, the correlation coefficients increase 0.004, 0.068, 0.05, 0.034 respectively for ML, GWR, RF and GRNN-based model.

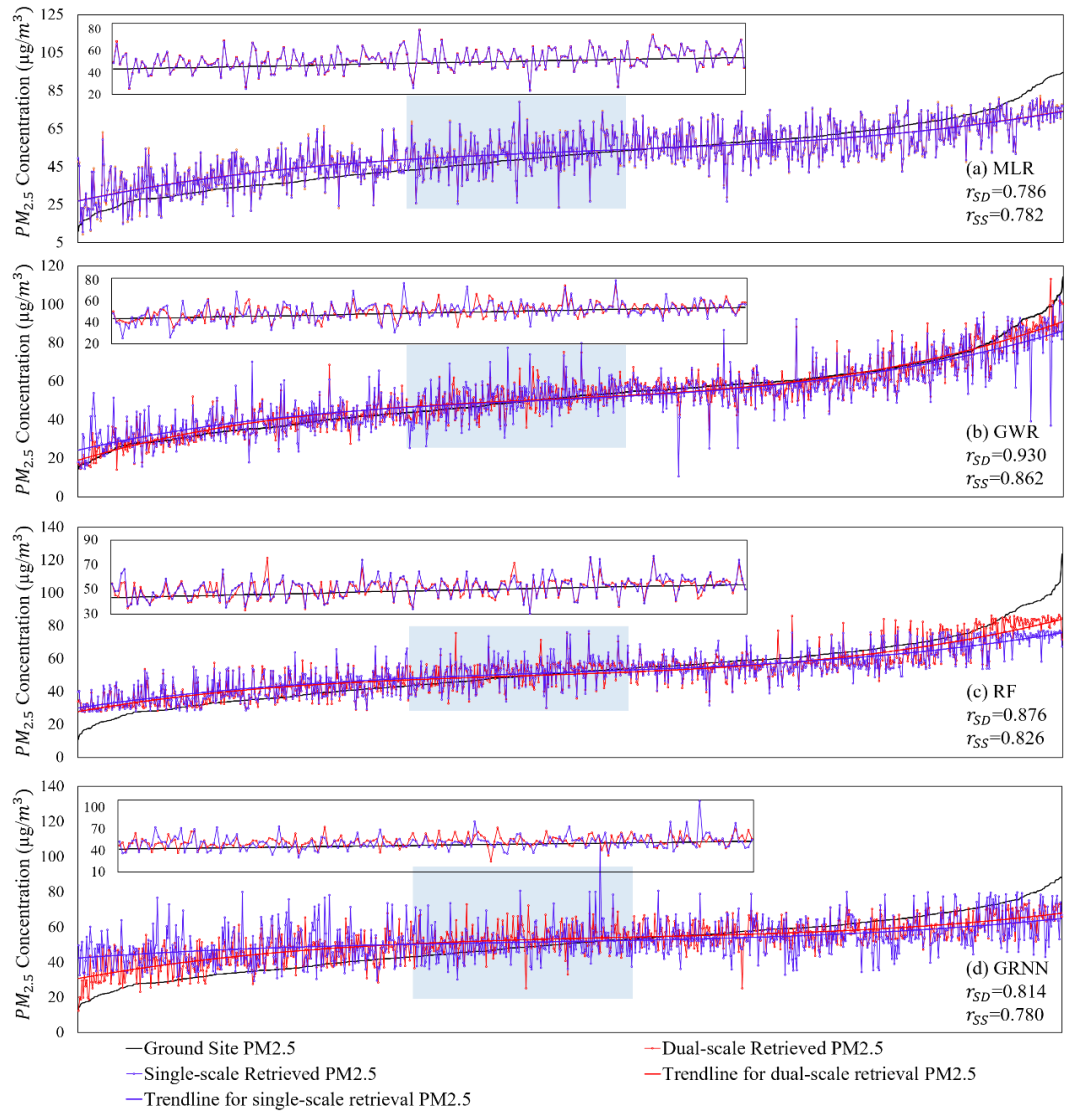

**Fig. 6** The dense point validation results of the four retrieval models. (a-d) stand for MLR, GWR, RF GRNN-based two scale retrieval models respectively. The sub figure in each figure is the enlarged view for the light blue part. The data are sort by site monitoring values from smallest to largest to make the comparisons with site values clearer. $r_{SD}/r_{SS}$ stands for the Pearson correlation coefficient between ground site PM2.5 and dual-scale/single-scale retrieved PM2.5 concentration.

*3.2. PM2.5 mapping at sub-km level*

The proposed dual-scale retrieval method can not only acquire higher prediction accuracy, and can also generate the final product with higher spatial resolution. We selected several typical cities in China and drew the PM2.5 concentration map at the resolution of 0.003°. For a comparison, the mapping results of both the single-scale and dual-scale retrieval are displayed in Fig. 7. We can clearly find that the PM2.5 mapping results of dual-scale retrieval has the same overall spatially varying trend as the results of single-scale retrieval. In addition, more detailed information can be captured by dual-scale retrieval than single-scale retrieval. For the five cities with different climates, topographies and pollution degrees, the proposed method shows good stability.

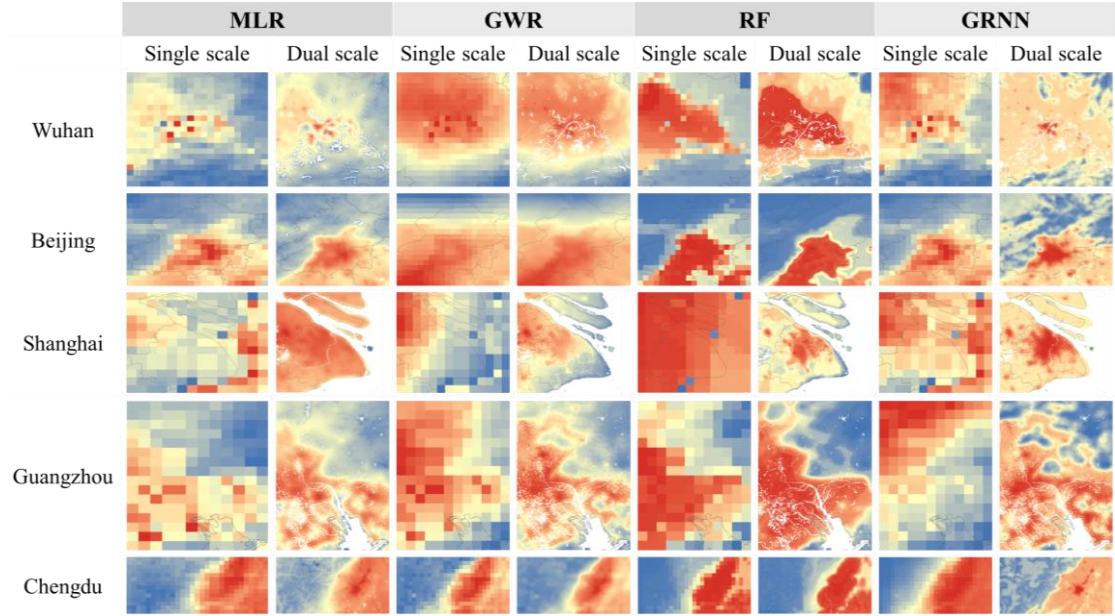

**Fig. 7.** The PM2.5 concentration distribution map in five typical cities using four different retrieval models. The results of both the single and dual-scale retrieval are displayed for a comparison. The white area in Dual-scale retrieval results are watershed.

Though the proposed method has a stronger spatial detail expression ability under all retrieval models, the mapping quality varies. RF has the worst performance among the four retrieval models, though with a quite satisfying performance on quantitative evaluation. The mapping results of RF shows blocking and stratification phenomena. This situation also occurred to some researchers in their work (Yuan et al., 2019). This may be caused by the intrinsic characteristic of RF algorithm. For the other three retrieval models, the PM2.5 concentration map is continuous in space and show a high mapping quality. Overall. The retrieval experiments in five different cities, with various pollution degree, climate and topography, proves that the dual-scale retrieval method is a robust retrieval method and can be applied to multifarious regions.

Furtherly, we selected the GWR-based dual-scale retrieval method for the mapping of PM2.5 concentration in Wuhan from 2013 to 2015 and the results are shown in Fig. 8. We can see that from 2013 to 2015, the PM2.5 concentration is decreasing overall. Specifically, for 2013 to 2014, the decreasing of PM2.5 concertation mostly happens in the suburbs, and from 2014 to 2015, the decreasing mainly concentrated in urban areas.

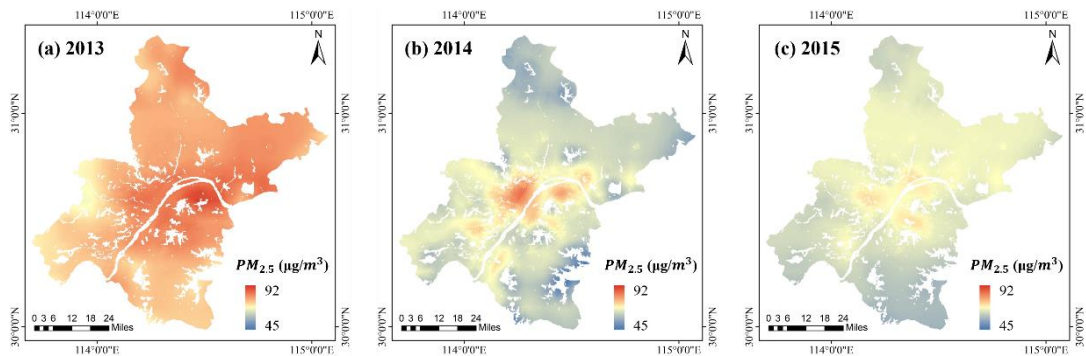

**Fig. 8.** Scatter plots for PM$_{2.5}$ concentration and AOD in the 9 metropolitan regions (The unit of PM$_{2.5}$ concentration is $\mu g/m^3$ ).

*3.3. Discussion*

*3.3.1 Exploration the performance difference of the four models*

Firstly, when conducting the retrieval using MLR model, the dual scale retrieval didn't improve much compared to the single-scale retrieval, and the cross-validation R$^2$ only increased 0.01 and the fitting $R^2$ didn't improve (Table 1). We think this may be explainable. A key point for dual-scale retrieval is that the information in the low-resolution variables (AOD, meteorological variables) was compressed in the retrieved low-resolution PM2.5 product, so in the second stage, the model mainly aims to describe the relationships between low and high resolution PM2.5 product, or to say, to describe the scale difference. We inferred that the relationships between low and high resolution PM2.5 is complex and is not a simple fixed linear relationship. So, MLR model in the second stage cannot make a full use of the information contained in the low-resolution PM2.5 product from the first stage, therefore, the loss of information makes the results don't improve much. To verify our conjecture, we made a supplementary experiment. We keep the retrieval model for the first stage being MLR model, and change the retrieval model of the second stage, to see whether the performance can be improved. The results turned out that, when the retrieval model for the second stage is not MLR but the GWR, RF and GRNN model, the performance can be promoted as shown in Table 2. This may prove that MLR model cannot describe the scale difference well, so didn't show much improvement in the dual-scale retrieval experiments.

Table 2. The performances for models with MLR as first-scale model and different models for second-scale.

|      | Fitting $R^2$ | CV $R^2$ |
|------|---------------|----------|
| MLR  | 0.64          | 0.63     |
| GWR  | 0.78          | 0.71     |
| RF   | 0.74          | 0.73     |
| GRNN | 0.66          | 0.65     |

Secondly, when conducting the retrieval using GWR model, overfitting appears in the first stage and single-scale retrieval. We infer that this is due to the multicollinearity problem. Meteorological factors can be correlated with each other in some cases, in first stage and single retrieval, all the nine meteorological factors are input which may bring about the multicollinearity problem and then result in the overfitting. In the second stage, the information contained in the meteorological factors was contracted in the retrieved PM2.5, so, the multicollinearity problem was alleviated and the overfitting disappeared in the second stage.

Thirdly, our results showed that GWR has the best performance, transcend the performance of ML-based models. That may be a little out of expectations for some readers. We think there are mainly two reasons for the explanation of this phenomena. First, it should be noticed that in our study the used data are annual data, and the temporal information was not considered, so no temporal predictions were made in our study. GWR is a model known for considering spatial heterogeneity, hence, it may perform statistically for spatial predictions. Second, the use of annual data makes the sample number used for training not large in the study, which is around 1430. The small amount of training samples limits the data mining ability of machine learning algorithms. Therefore, GWR can perform better than ML-based algorithms. This remind us that though ML-algorithms can achieve pretty good performance in many cases, it doesn't apply for all situations. For example, as a data-driven algorithm, ML may not be suitable for studies without massive data. And some traditional models have more potential to be mined. The combination of ML and geographical or geostatistical knowledge may worth more effort.

Finally, RF model also shows a good performance in terms of quantitative evaluation, however, the mapping results can be a little "strange" in some cases. For example, in the results of Beijing , we can see stratification; in results of Wuhan, there are some patches, all of which makes the mapping results not continuous and smooth enough in space. Therefore, though with a high cross validation score, RF still performances bad in mapping results. This remind us that when evaluate the performance of a retrieval model, quantitative indicator is not enough, the mapping performance is important as well.

*3.3.2 Limitations and future work*

In the mapping result of dual-scale retrieval, we can sometimes see the trace of DEM and landcover map. For example, in Fig.7, the left part of the MLR and GRNN retrieved results in Chengdu looks like texture of DEM. And GRNN retrieved results in Shanghai can looks similar to the texture of landcover. We think is due to the limited input variables in the second stage. In the second stage, among the input variables, there are only two with true and valid detail information: DEM and landcover. The provided information may be very limited, thus making the retrieval results looks like the texture of the input variables in some cases. In the future, we would like to explore and introduce more variables which have high resolution and have impact on PM2.5 concertation. The introduce of multisource high-resolution data may be able to describe the detailed variations of PM2.5 concentration as it is in a better way.

In this study, we only test the performance of dual-scale retrieval, retrieval using more scales are not tried for fear that the repeat resampling of low-resolution product may bring large uncertainties thus decrease the model accuracy and generalization ability. Besides, the necessity of retrieving PM2.5 at 0.0003° (~30m) is not that obvious. In the future, if needed, and if there are enough data at multiple resolutions, it worth a try to expanded the dual-scale retrieval to multi-scale retrieval.

**4. Conclusions**

Traditional satellite-based PM2.5 retrieval method achieves the PM2.5 mapping at the resolution of AOD with all the auxiliary variables resampled to the resolution of AOD, regardless of the fact that variables with higher resolution than AOD may contain important detail information for capturing spatial variations of PM2.5 at fine scale. In this study, we propose a dual-scale retrieval method to make a fuller use of information contained in the variables with different resolution. Variables with low resolution are used for the first-scale retrieval at coarse scale, and then variables at higher resolution are used for the retrieval at fine scale. As a connection between the retrieval at the two scales, PM2.5 product of the first stage is upsampled and input into the model at second stage. The results of four retrieval models, i.e. MLR, GWR, RF, GRNN, show that dual-scale retrieval can achieves a higher estimation accuracy, and map the PM2.5 concentration with more correct detail than single-scale retrieval. Among the four models, GWR shows the best performance considering both quantitative evaluation and mapping quality. Therefore, a GWR-based dual-scale retrieval for PM2.5 concentration at Wuhan was conducted for 2013 to 2015, the spatial and annual variations are analyzed at fine scale.

**Acknowledgments**

This work was supported by the National Key R&D Program of China (no. 2016YFC0200900). We would like to thank the $PM_{2.5}$ data providers of the China National Environmental Monitoring Center (CNEMC). The MODIS AOD were obtained from the NASA Langley Research Center Atmospheric Science Data center (ASDC). The NCEP/NCAR Reanalysis data are provided by the NOAA/OAR/ESRL PSD. The MERRA-2 data A were provided by the NASA GES DISC. Finally, we also thank the ESA team for providing the landcover data.

**Conflicts of Interest**

The authors declare no conflict of interest.

References

**Uncategorized References**

Beloconi, A., Kamarianakis, Y., Chrysoulakis, N., 2016. Estimating urban PM10 and PM2.5 concentrations, based on synergistic MERIS/AATSR aerosol observations, land cover and morphology data. Remote Sensing of Environment 172, 148-164.

Bi, J., Belle, J.H., Wang, Y., Lyapustin, A.I., Wildani, A., Liu, Y., 2019. Impacts of snow and cloud covers on satellite-derived PM2.5 levels. Remote Sensing of Environment 221, 665-674.

Boys, B.L., Martin, R.V., van Donkelaar, A., MacDonell, R.J., Hsu, N.C., Cooper, M.J., Yantosca, R.M., Lu, Z., Streets, D.G., Zhang, Q., Wang, S.W., 2014. Fifteen-year global time series of satellite-derived fine particulate matter. Environmental science & technology 48, 11109-11118.

Chen, G., Li, S., Knibbs, L.D., Hamm, N.A.S., Cao, W., Li, T., Guo, J., Ren, H., Abramson, M.J., Guo, Y., 2018. A machine learning method to estimate PM2.5 concentrations across China with remote sensing, meteorological and land use information. The Science of the total environment 636, 52-60.

Guo, Y., Tang, Q., Gong, D.-Y., Zhang, Z., 2017. Estimating ground-level PM 2.5 concentrations in Beijing using a satellite-based geographically and temporally weighted regression model. Remote Sensing of Environment 198, 140-149.

He, Q., Huang, B., 2018. Satellite-based mapping of daily high-resolution ground PM 2.5 in China via space-time regression modeling. Remote Sensing of Environment 206, 72-83.

Ho, H.C., Wong, M.S., Yang, L., Chan, T.C., Bilal, M., 2018. Influences of socioeconomic vulnerability and intra-urban air pollution exposure on short-term mortality during extreme dust events. Environmental pollution 235, 155-162.

Hu, X., Belle, J.H., Meng, X., Wildani, A., Waller, L.A., Strickland, M.J., Liu, Y., 2017. Estimating PM2.5 Concentrations in the Conterminous United States Using the Random Forest Approach. Environmental science & technology 51, 6936-6944.

Hu, X., Waller, L.A., Al-Hamdan, M.Z., Crosson, W.L., Estes, M.G., Jr., Estes, S.M., Quattrochi, D.A., Sarnat, J.A., Liu, Y., 2013. Estimating ground-level PM(2.5) concentrations in the southeastern U.S. using geographically weighted regression. Environmental research 121, 1-10.

Huang, K., Xiao, Q., Meng, X., Geng, G., Wang, Y., Lyapustin, A., Gu, D., Liu, Y., 2018. Predicting monthly high-resolution PM2.5 concentrations with random forest model in the North China Plain. Environmental pollution 242, 675-683.

Jiang, M., Sun, W., Yang, G., Zhang, D., 2017. Modelling Seasonal GWR of Daily PM2.5 with